\documentclass[a4paper,10pt]{article}
\usepackage[T1]{fontenc}
\usepackage[utf8x]{inputenc}
\usepackage{amsmath}
\usepackage{amsfonts}
\usepackage{natbib}
\usepackage{amsthm}

\title{Noise, risk premium, and bubble}
\author{Grzegorz Andruszkiewicz \and Dorje C. Brody}

\begin{document}
\maketitle

\newcommand{\half}{\mbox{$\textstyle \frac{1}{2}$}}
\newcommand{\quat}{\mbox{$\textstyle \frac{1}{4}$}}
\newcommand{\re}{\mbox{$\rm e$}}
\newcommand{\rd}{\mbox{$\rm d$}}

\begin{center}
Department of Mathematics, Imperial College London,
London SW7 2AZ, UK
\end{center}

\begin{abstract}
The existence of the pricing kernel is shown to imply the existence of an 
ambient information process that generates market filtration. This information 
process consists of a signal component concerning the value of the random 
variable $X$ that can be interpreted as the timing of future cash demand, 
and an independent noise component. The conditional expectation of the 
signal, in particular, determines the market risk premium vector. An addition to 
the signal of any term that is independent of $X$, which generates a drift in 
the noise, is shown to change the drifts of price processes in the physical 
measure, without affecting the current asset price levels. Such a drift in the 
noise term can induce anomalous price dynamics, and can be seen to explain 
the mechanism of observed phenomena of equity premium and financial bubbles. 
\end{abstract}

\section{Introduction}

The market risk premium is one of the main factors that drives the return of 
any given portfolio of assets. Hence it is a key quantity for hedge funds, pension 
funds, and numerous other investors. The risk premium can make investments 
grow smoothly or jump up and down widely, often in an unpredictable manner. 
In spite of its importance in asset allocation, however, the risk premium is 
notoriously difficult to estimate from observed price processes of various risky 
assets (see, e.g., Rogers 2001). Is it possible then to estimate the risk premium 
from current prices of financial derivatives? 

If $\{S_t\}$ denotes the price process of a risky asset and $h(s)$ is the payout 
function of a European contingent claim expiring at $T$, then the price of this 
derivative is given by the expectation of the cash flow $h(S_T)$, suitably 
discounted, in the risk-neutral measure. Because asset price processes in 
the risk-neutral measure are independent of the market risk premium, one 
might be tempted to conclude therefore that derivative prices are likewise 
independent of the risk premium. Indeed, in the case of the 
Black-Scholes-Merton model where all relevant parameters are constant in 
time, the risk premium parameter essentially drops out of various derivative 
pricing formulae. Notwithstanding this example, it is worth bearing in mind that 
the choice of the pricing measure does depend on the choice of the risk 
premium. Thus, derivative prices in general will depend implicitly on the risk 
premium, often in a nonlinear way. It follows that calibration of the market 
risk premium from option prices is feasible within a given modelling framework 
(Brody \textit{et al}. 2011). 

The main purpose of the present paper is to address the question whether it is 
possible, at least in principle, to determine the risk premium unambiguously, if 
the totality of arbitrage-free market prices for various derivatives were available. 
We shall find that the market risk premium consists of two components in an 
additive manner (for models based on Brownian filtrations): The first of 
the two, which we might call a `systematic' component, depends explicitly on 
the term structure of the market, while the second, which we might call an 
`idiosyncratic' component, is independent of the term structure of the market, 
and thus can be identified as \textit{pure noise}. We show that the systematic 
component can in principle be determined from current market data, whereas 
the idiosyncratic noise component is strictly `hidden' and thus cannot be 
inferred from derivative prices. Therefore, the risk premium can be backed 
out from market data only up to an indeterminable additive noise. 

Although the noise component cannot be inferred directly, it nevertheless has 
an impact on the dynamics of asset prices under the physical measure, even 
though it does not reflect the `true' state of affairs. Hence a spontaneous creation 
of superfluous noise can move the price of an asset in an essentially arbitrary 
direction. In particular, because the risk premium, and hence its noise component, 
is a vectorial quantity, the direction of the noise vector can at times lie close to 
the directions of volatility vectors of the share prices of a particular industrial sector, 
leading to the creation of a `bubble' for that sector by pushing up those share 
prices. When a more reliable information concerning the state of that sector is 
unveiled, the direction of the risk premium vector is likely to change so as to 
generate a negative component in the excess rate of return. This can be 
exacerbated by an increase in the magnitude of asset volatilities due to 
information revelation, 
thus leading to a `burst'. Such a scenario need not be confined to a particular 
financial sector; the existence of the so-called `equity premium puzzle' over a 
specified period can likewise be attributed to the prevailing noise that points in 
the general direction of the equity market volatility, but not in the direction of the 
bond market volatility. 

Needless to say, our formulation does not explain the cause of the creation 
of anomalous price movements such as a financial bubble or an equity premium; 
nor does it address the predictability of these events. In fact, according to our 
characterisation, bubbles can at best be identified retrospectively, after their 
bursts. Nevertheless, we are able to describe the mechanism by which such 
anomalous price movements are generated in a simple and intuitive manner. In 
particular, since our characterisation of a bubble is different from those more 
commonly used in the literature, we are able to circumvent the analysis based on 
subtle distinctions between local and true martingales. We also provide a heuristic 
argument why the hidden noise might have the tendency of creating equity premium.

\section{Pricing kernel}

For definiteness, we shall be adopting the pricing kernel approach (see, e.g., 
Cochrane 2005, Bj\"ork 2009). We model the financial market on a probability 
space $(\Omega,\mathcal{F}, \mathbb{P})$ with filtration 
$\{{\mathcal F}_t\}_{t\geq0}$. Here 
$\mathbb{P}$ denotes the `physical' probability measure, and  
$\{{\mathcal F}_t\}$ is assumed to be generated by a multi-dimensional 
Brownian motion. Expectation under 
${\mathbb P}$ is denoted $\mathbb{E}[-]$, and for the conditional expectation 
with respect to ${\mathcal F}_t$ we write ${\mathbb E}_t[-]$. Two other probability 
measures enter the ensuring discussion; these are the risk-neutral measure 
${\mathbb Q}$ and an auxiliary measure ${\mathbb R}$ to be described below. 
Expectations in these measures will be written $\mathbb{E}^{\mathbb Q}[-]$ and 
$\mathbb{E}^{\mathbb R}[-]$, respectively. 

We assume that the market is free of arbitrage opportunities, and that there is 
an established pricing kernel, but market completeness is not assumed. These 
assumptions imply the existence of a unique preferred risk-neutral measure 
${\mathbb Q}$. The pricing kernel, denoted here by $\{\pi_t\}_{t\geq0}$, is a 
positive supermartingale with the property that if $S_T$ is the price at time $T$ 
of an asset that pays no dividend, then the price at time $t$ of the asset is given 
by 
\begin{eqnarray}
S_t = \frac{1}{\pi_t}\, {\mathbb E}_t [\pi_T S_T] .
\label{eq:1}
\end{eqnarray}
In particular, if $S_T=1$, then (\ref{eq:1}) gives the pricing formula for the discount 
bond: $P_{tT} = {\mathbb E}_t [\pi_T]/\pi_t$. 

We shall proceed by discussing some properties of the pricing kernel that are 
relevant to our analysis here. In addition to being a positive supermartingale, the 
pricing kernel fulfils the condition that ${\mathbb E}[\pi_t]\to0$ as $t\to\infty$. A 
positive supermartingale possessing this property is known as a \textit{potential}. 
It follows that every pricing kernel can be represented as a potential, and conversely 
every potential constitutes an admissible pricing kernel. The Doob-Meyer 
decomposition then shows that $\{\pi_t\}$ can be represented uniquely in the form: 
\begin{eqnarray}
\pi_t = {\mathbb E}_t[A_\infty] - A_t,
\end{eqnarray}
where $\{A_t\}_{t\geq0}$ is an increasing adapted process such that $A_\infty$ is 
finite. Note that ${\mathbb E}_t[A_\infty]$ is a uniformly integrable martingale. We 
may define $\{A_t\}$ according to 
\begin{eqnarray}
A_t = \int_0^t a_s \rd s
\end{eqnarray}
for some adapted nonnegative process $\{a_t\}$. Hence it suffices to choose the 
process $\{a_t\}$ to model the pricing kernel, and this leads to the potential 
approach of Rogers (1997) to model term structure dynamics. A substitution shows 
that 
\begin{eqnarray}
\pi_t = \int_t^{\infty} {\mathbb E}_t [a_u] \rd u.
\label{eq:4}
\end{eqnarray}
The representation (\ref{eq:4}) resembles that of Flesaker and Hughston (1996,1997), 
if we make the following identification. First, writing $\rho_0(T) = -\partial_T P_{0T}$, 
where $P_{0T}$ is the initial discount function, we see that the processes 
$\{M_t(u)\}_{t\geq0,u\geq t}$ defined by 
\begin{eqnarray}
M_{t}(u) = \frac{{\mathbb E}_t[a_u]}{\rho_0(u)}
\end{eqnarray}
is a one-parameter family of positive martingales, i.e. for each fixed $u\geq t$, 
$\{M_t(u)\}$ is a martingale. This follows on account of the martingale property 
of the conditional expectation ${\mathbb E}_t[a_u]$. In terms of these positive 
martingales, the pricing kernel can be expressed in the Flesaker-Hughston 
form:
\begin{eqnarray}
\pi_t = \int_t^{\infty} \rho_0(u) M_t(u) \rd u.
\label{eq:6}
\end{eqnarray}
From the martingale representation theorem we deduce that the dynamical equations 
satisfied by the positive martingale family $\{M_t(u)\}$ take the form: 
\begin{eqnarray}
\rd M_t(u) = M_t(u) v_t(u) \rd \xi_t,
\label{eq:7}
\end{eqnarray}
where $\{v_t(u)\}$ is a family of adapted (in general vectorial) processes and 
$\{\xi_t\}$ is a standard multi-dimensional 
Brownian motion under the ${\mathbb P}$ measure. We observe therefore that 
modelling the pricing kernel is equivalent to modelling the one-parameter family 
of volatility processes $\{v_t(u)\}$. On account of (\ref{eq:6}) and (\ref{eq:7}) we 
deduce, by an application of Ito's lemma, that 
\begin{eqnarray}
\frac{\rd \pi_t}{\pi_t} = -r_t \rd t - \lambda_t  \rd \xi_t,
\label{eq:dpi}
\end{eqnarray}
where 
\begin{eqnarray}
r_t = \frac{\rho_0(t) M_t(t)}{\int_t^{\infty} \rho_0(u) M_t(u) \rd u}
\end{eqnarray}
is the short rate, and 
\begin{eqnarray}
\lambda_t = -\frac{\int_t^\infty \rho_0(u) v_t(u) M_t(u)\rd u}
{\int_t^{\infty} \rho_0(u) M_t(u) \rd u}
\label{eq:10}
\end{eqnarray}
is the market risk premium. The fact that the drift 
of $\{\pi_t\}$ can be identified with the short rate can be seen by applying the 
martingale condition (\ref{eq:1}) on the money market account $\{B_t\}$ 
satisfying $\rd B_t = r_t B_t \rd t$. That is, the drift of $\{\pi_t B_t\}$ vanishes if 
and only if the drift of $\{\pi_t\}$ is $\{-r_t\}$. Similarly, let us write $\{\mu_t\}$ for 
the drift of a risky asset $\{S_t\}$ that pays no dividend, and $\{-\lambda_t\}$ for 
the volatility of $\{\pi_t\}$. Then the martingale condition on $\{\pi_t S_t\}$ implies 
that $\mu_t=r_t+\lambda_t\sigma_t$, which shows that $\{\lambda_t\}$ indeed 
expresses the excess rate of return above the risk-free rate in unit of volatility. 

An advantage of working with the pricing kernel is that once a model is chosen 
for the volatility processes $\{v_t(u)\}$ of the martingale family, we are able not 
only to price a wide range of derivatives via the pricing formula 
${\mathbb E}[\pi_T H_T]$, where $H_T$ is the payout of a derivative, but also 
to obtain a model for the interest rate term structure. Furthermore, a model for 
$\{v_t(u)\}$, which can be calibrated by use of market data for derivative prices, 
implies a process for the risk premium $\{\lambda_t\}$ according to the 
prescription (\ref{eq:10}), and this in turn can be used for asset allocation purposes. 
This is the sense in 
which derivative prices can be used to calibrate the risk premium, within any 
modelling framework (cf. Brody \textit{et al}. 2011). The issues that we would 
like to address here are: (a) the ambiguity associated with the determination of 
the risk premium from market data; and (b) the identification of the origin of this 
ambiguity. For these purposes, it is useful to examine the probabilistic 
characterisation of the pricing kernel, within the term structure density approach 
of Brody and Hughston (2001).

\section{Probabilistic representation of the pricing kernel}

To proceed, we shall make the following observation that the positivity of 
nominal interest and the requirement that a bond with infinite maturity must 
have vanishing value imply that $\rho_0(T) = -\partial_T P_{0T}$ defines a 
probability density function on the positive half-line (Brody and Hughston 2001). 
More generally, the positivity of the martingale family $\{M_t(u)\}$ implies that 
$\{\rho_t(u)\}$ defined by 
\begin{eqnarray}
\rho_t(u) = \frac{\rho_0(u) M_t(u)}{\int_0^{\infty} \rho_0(u) M_t(u) \rd u}
\label{eq:12}
\end{eqnarray}
is a measure-valued process, i.e. $\rho_t(u)\geq 0$ for all $t$ and all $u$; and 
\begin{eqnarray}
\int_0^\infty \rho_t(u) \rd u = 1 
\end{eqnarray}
for all $t\geq0$. 
The measure-valued process thus introduced suggests the existence of a random 
variable $X$ whose conditional density under some probability measure is given by 
(\ref{eq:12}). Furthermore, an application of Ito's lemma on (\ref{eq:12}) shows that 
\begin{eqnarray}
\frac{\rd \rho_t(u)}{\rho_t(u)} = \left( v_t(u) - {\hat v}_t\right) 
\left( \rd\xi_t - {\hat v}_t\rd t\right), 
\label{eq:14} 
\end{eqnarray}
where 
\begin{eqnarray}
{\hat v}_t = \int_0^\infty v_t(u) \rho_t(u) \rd u
\end{eqnarray}
can be thought of as the conditional expectation of $v_t(X)$. 

Indeed, the dynamical equation (\ref{eq:14}) takes the form of a Kushner equation, 
thus implies the existence of the following auxiliary filtering problem. For simplicity 
of exposition, let us for now assume that the one-parameter family of volatility 
processes $\{v_t(u)\}$ is deterministic. We introduce a probability space 
$(\Omega,{\mathcal F},{\mathbb R})$, upon which $X$ is a positive random variable 
with density $\rho_0(u)$. The meaning of the measure ${\mathbb R}$ will be 
examined at a later point. On this probability space, consider the following 
information process 
\begin{eqnarray}
\xi_t = \int_0^t v_s(X) \rd s + \beta_t , 
\label{eq:16}
\end{eqnarray}
where $\{\beta_t\}$ is an ${\mathbb R}$-Brownian motion, independent of $X$. The 
task of the `observer' is thus to determine the best estimate of $X$ given the data 
$\{\xi_s\}_{0\leq s\leq t}$. Assuming that the criteria for optimality is an estimator that 
minimises the quadratic error, standard results in filtering theory (cf. Wonham 1965, 
Liptser and Shiryaev 2001) show that the best estimate for $X$ is the expectation 
of $X$ with respect to the \textit{a posteriori} density: 
\begin{eqnarray}
\frac{\rd}{\rd u} {\mathbb R}(X<u|{\mathcal F}_t) = \frac{\rho_0(u) 
\exp\left(\int_0^t v_s(u){\rm d}\xi_s - \frac{1}{2}\int_0^t v_s^2(u){\rm d}s\right)}
{\int_0^{\infty} \rho_0(u) \exp\left(\int_0^t v_s(u){\rm d}\xi_s - \frac{1}{2}\int_0^t 
v_s^2(u){\rm d}s\right) \rd u},
\label{eq:17}
\end{eqnarray}
where ${\mathcal F}_t=\sigma(\{\xi_s\}_{0\leq s\leq t})$.
Notice that the right side of (\ref{eq:17}) is in fact identical to the right side of 
(\ref{eq:12}), provided that the measure change between ${\mathbb P}$ and 
${\mathbb R}$ are suitably defined (recall that $\{\xi_t\}$ in the ${\mathbb P}$ 
measure is a Brownian motion, while in the ${\mathbb R}$ measure it is a 
drifted Brownian motion of (\ref{eq:16})). 

The above-specified filtering problem leads to the following probabilistic interpretation 
for the pricing kernel. Writing 
\begin{eqnarray}
N_t = \int_0^{\infty} \rho_0(u) \exp\left(\int_0^t v_s(u){\rm d}\xi_s - \half
\int_0^t v_s^2(u){\rm d}s\right) \rd u
\label{eq:18}
\end{eqnarray}
for the normalisation of the conditional density $\{\rho_t(u)\}$, we see that the 
pricing kernel is given by the `unnormalised' conditional probability that $X>t$: 
\begin{eqnarray}
\pi_t = N_t \, {\mathbb R}_t(X>t),
\label{eq:19} 
\end{eqnarray}
where for simplicity we have written ${\mathbb R}_t(-)={\mathbb R}(-|{\mathcal F}_t)$ 
for the conditional probability. Further, the price of a discount bond admits a 
probabilistic representation in the ${\mathbb R}$ measure: 
\begin{eqnarray}
P_{tT} = \frac{{\mathbb R}_t(X>T)}{{\mathbb R}_t(X>t)}.
\label{eq:20} 
\end{eqnarray}
This formula shows that the price process of a discount bond is given by the 
ratio of the ${\mathbb R}$-conditional probability that the positive random variable 
$X$ taking values greater than $T$ and that of $X$ taking values greater than $t$. 
By use of the Bayes formula, we deduce that (\ref{eq:20}) can alternatively be 
expressed in the form 
\begin{eqnarray}
P_{tT} = {\mathbb R}_t(X>T|X>t) ,
\label{eq:21} 
\end{eqnarray}
since the set $\{X>t\}$ contains the set $\{X>T\}$. This is essentially the 
representation obtained by Brody and Friedman (2009) for the discount bond 
using the information-based approach to interest rate modelling. 

It is worth remarking that the random variable $X$ has the dimension of time. 
In Brody and Friedman (2009), $X$ was interpreted as the arrival time of liquidity 
crisis, in the narrow sense of a cash demand. Hence, under this interpretation, 
(\ref{eq:21}) shows that the bond price at $t$ is the probability that the timing of 
the occurrence of a cash demand is beyond $T$, given that it has not yet 
occurred at $t$, and given the noisy information (\ref{eq:16}) concerning the value 
of $X$, in a suitably chosen measure ${\mathbb R}$. 

\section{Back to the market measure}

The normalisation $\{N_t\}$ can be used to effect a measure change ${\mathbb R} 
\to {\mathbb P}$. To see this, note first that the process $\{W_t\}$ defined by 
\begin{eqnarray}
W_t = \xi_t - \int_0^t {\mathbb E}_s^{\mathbb R}[v_s(X)] \rd s 
\label{eq:22}
\end{eqnarray}
is an ${\mathbb R}$-Brownian motion with respect to the filtration $\{{\mathcal F}_t\}$ 
generated by the information process (\ref{eq:16}). In fact, this is just the innovations 
representation for the filtering problem posed above. Thus, the Brownian property 
can be verified by checking that $\{W_t\}$ satisfies the martingale condition: 
\begin{eqnarray}
{\mathbb E}_t^{\mathbb R}\left[W_T\right] &=& {\mathbb E}_t^{\mathbb R}\left[ 
\int_0^T v_s(X) \rd s + \beta_T - \int_0^T {\mathbb E}_s^{\mathbb R}[v_s(X)] \rd s 
\right] \nonumber \\ &=& 
{\mathbb E}_t^{\mathbb R}\left[ \int_0^t v_s(X) \rd s + \beta_t - \int_0^t 
{\mathbb E}_s^{\mathbb R}[v_s(X)] 
\rd s \right] \nonumber \\ && +  {\mathbb E}_t^{\mathbb R}\left[ \int_t^T v_s(X) \rd s 
+ (\beta_T-\beta_t) - \int_t^T {\mathbb E}_s^{\mathbb R}[v_s(X)] \rd s \right] 
\nonumber \\ &=& W_t,
\end{eqnarray}
where we have made use of the martingale property 
${\mathbb E}_t^{\mathbb R}[{\mathbb E}_s^{\mathbb R}[v_s(X)]]=
{\mathbb E}_t^{\mathbb R}[v_s(X)]$ of the conditional expectation for $s>t$, 
and the tower property of 
conditional expectation to deduce ${\mathbb E}_t^{\mathbb R}[\beta_T] = 
{\mathbb E}_t^{\mathbb R}[\beta_t]$. Along with $(\rd W_t)^2=\rd t$, L\'evy's 
characterisation shows that $\{W_t\}$ is an ${\mathbb R}$-Brownian motion. 

On the other hand, an application of Ito's lemma on (\ref{eq:18}) gives 
\begin{eqnarray}
\frac{\rd N_t}{N_t} = {\hat v}_{t} \rd \xi_t,
\end{eqnarray}
from which it follows that 
\begin{eqnarray}
N_t = \exp \left( \int_0^t {\hat v}_{s} \rd \xi_s - \half \int_0^t {\hat v}_{s}^2 
\rd s \right) 
\label{eq:25}
\end{eqnarray}
is a positive martingale satisfying $N_0=1$. Hence $\{N_t\}$ can be used as the 
likelihood process to change the probability measure. Specifically, for any 
${\mathcal F}_t$-measurable random variable $Z_t$ we have 
\begin{eqnarray}
{\mathbb E}_s^{\mathbb R}\left[Z_t\right] = \frac{1}{N_s}{\mathbb E}_t^{\mathbb P}
\left[N_tZ_t\right] \quad {\rm and} \quad {\mathbb E}_s^{\mathbb P}\left[Z_t\right] = 
N_s {\mathbb E}_t^{\mathbb R}\left[\frac{1}{N_t}Z_t\right].
\end{eqnarray}
In particular, (\ref{eq:22}) and (\ref{eq:25}) shows that $\{\xi_t\}$ is a Brownian 
motion under the ${\mathbb P}$ measure. In addition, we find that the random 
variable $X$ has the same probability law under ${\mathbb P}$ as under 
${\mathbb R}$, and that $X$ and $\xi_t$ are ${\mathbb P}$-independent. These 
properties can be verified by showing 
\begin{eqnarray}
{\mathbb E}^{\mathbb P}[\re^{{\rm i}(x\xi_t+yX)}]=
{\mathbb E}^{\mathbb P}[\re^{{\rm i}x\xi_t}]{\mathbb E}^{\mathbb P}[\re^{{\rm i}yX}]  
\end{eqnarray}
for all real $x,y$, and calculating the right side explicitly. 

We remark that the conditional probability ${\mathbb R}_t(X>t)$ appearing in 
(\ref{eq:19}) can be interpreted as representing the pricing kernel in the 
${\mathbb R}$ measure. Specifically, writing $\Pi_t$ for ${\mathbb R}_t(X>t)$, 
we deduce from (\ref{eq:17}) that 
\begin{eqnarray}
\Pi_t = \frac{\int_t^\infty \rho_0(u) \exp\left( 
\int_0^t v_s(u){\rm d}\xi_s - \frac{1}{2}\int_0^t v_s^2(u){\rm d}s\right)\rd u}
{\int_0^{\infty} \rho_0(u) \exp\left(\int_0^t v_s(u){\rm d}\xi_s - \frac{1}{2}\int_0^t 
v_s^2(u){\rm d}s\right) \rd u}.
\label{eq:27}
\end{eqnarray}
A short calculation making use of (\ref{eq:22}) shows that the 
${\mathbb R}$-pricing kernel (\ref{eq:27}) can be expressed manifestly in the 
Flesaker-Hughston representation: 
\begin{eqnarray}
\Pi_t = \int_t^\infty \rho_0(x) G_t(x) \rd x,
\end{eqnarray}
where $\{G_t(x)\}$ is a one-parameter family of positive $\mathbb{R}$-martingales:
\begin{eqnarray}
G_t(x) = \exp \left( \int_0^t \tilde{v}_s(x) \rd W_t -\half \int_0^t \tilde{v}_s(x)^2 
\rd t \right), 
\label{eq:29}
\end{eqnarray}
and where $\tilde{v}_t(x) = v_t(x) - {\mathbb E}_t^{\mathbb R}[v_t(X)]$. The 
dynamical equation satisfied by the $\mathbb{R}$-pricing kernel therefore reads 
\begin{eqnarray}
\frac{\rd \Pi_t}{\Pi_t} = -r_t \rd t - ({\hat v}_{t} +\lambda_t) \rd W_t,
\end{eqnarray}
where ${\hat v}_{t} = {\mathbb E}_t^{\mathbb R}[v_t(X)]$ and 
$\lambda_{t} = -{\mathbb E}_t^{\mathbb R}[v_t(X)|X>t]$.

\section{Indeterminacy of the risk premium} 

Returning to the ${\mathbb P}$-measure, we recall that once a parametric model 
for the martingale volatility $\{v_t(x)\}$ is chosen, then prices of derivatives will 
in general depend on this model choice. Hence $\{v_t(x)\}$ can be calibrated from 
derivative prices. The initial term structure density $\rho_0(u)$, on the other hand, 
can be calibrated from the initial yield curve. By substituting these ingredients 
in (\ref{eq:10}) we thus obtain a market implied risk premium, subject to the model 
choice. Of course, any tractable model is unlikely to fit all derivative prices. One 
can nevertheless ask whether it is possible to fix $\{v_t(x)\}$ in a hypothetical 
situation where one has access to the totality of liquidly-traded derivative prices 
and an unlimited computational resource, i.e. whether it is possible in principle to 
fix $\{v_t(x)\}$ unambiguously. Perhaps not surprisingly, the answer is negative. 

To see this, suppose that the volatility of the Flesaker-Hughston martingale family 
is decomposed in the form 
\begin{eqnarray}
v_t(u) = \phi_t(u) - \alpha_t , 
\label{eq:31}
\end{eqnarray}
where the vector process $\{\alpha_t\}$ is independent of $X$, and has no 
parametric dependence on $u$. The minus sign here is purely a matter of 
convention. Then writing 
\begin{eqnarray}
L_t = \exp\left(-\int_0^t \alpha_s \rd \xi_s - \half \int_0^t \alpha^2_s \rd s\right),
\end{eqnarray}
we find that the pricing kernel takes the form
\begin{eqnarray}
\pi_t = L_t \int_t^\infty \rho_0(u)\, \re^{ \int_0^t \phi_s(u) {\rm d} \xi_s - 
\frac{1}{2} \int_0^t \phi^2_s(u) {\rm d} s + \int_0^t \phi_s(u)\alpha_s{\rm d} s} \rd u.
\end{eqnarray}
It follows that the price at time $t$ of a contingent claim, with payout $H_T=
h(S_T)$ at $T>t$, is given by 
\begin{eqnarray}
H_t &=& {\mathbb E}_t\left[ \frac{\pi_T}{\pi_t}\, H_T \right] \nonumber \\ &=& 
{\mathbb E}_t\left[ \frac{L_T}{L_t} \frac{\int_T^\infty \rho_0(u)\, \re^{ \int_0^T 
\phi_s(u) {\rm d} \xi_s - \frac{1}{2} \int_0^T \phi^2_s(u) {\rm d} s + \int_0^T 
\phi_s(u)\alpha_s{\rm d} s} \rd u}{\int_t^\infty \rho_0(u)\, \re^{ \int_0^t \phi_s(u) 
{\rm d} \xi_s - \frac{1}{2} \int_0^t \phi^2_s(u) {\rm d} s + \int_0^t \phi_s(u) 
\alpha_s{\rm d} s} {\rm d} u} \, H_T \right] \nonumber \\ &=& 
{\mathbb E}_t^\alpha\left[ \frac{\int_T^\infty \rho_0(u)\, \re^{ \int_0^T 
\phi_s(u) {\rm d} \xi_s - \frac{1}{2} \int_0^T \phi^2_s(u) {\rm d} s + \int_0^T 
\phi_s(u)\alpha_s{\rm d} s} \rd u}{\int_t^\infty \rho_0(u)\, \re^{ \int_0^t \phi_s(u) 
{\rm d} \xi_s - \frac{1}{2} \int_0^t \phi^2_s(u) {\rm d} s + \int_0^t \phi_s(u) 
\alpha_s{\rm d} s} {\rm d} u} \, H_T \right], 
\label{eq:34}
\end{eqnarray}
where we have used $\{L_t\}$ as a density martingale to change the measure. 
Evidently, under the new measure ${\mathbb P}^\alpha$, the process 
$\{\xi_t^\alpha\}$ defined by 
\begin{eqnarray}
\xi_t^\alpha = \xi_t + \int_0^t \alpha_s \rd s
\label{eq:35}
\end{eqnarray}
is a standard Brownian motion. Substituting (\ref{eq:35}) in (\ref{eq:34}) we deduce 
that  
\begin{eqnarray}
H_t = 
{\mathbb E}_t^\alpha\left[ \frac{\int_T^\infty \rho_0(u)\, \re^{ \int_0^T 
\phi_s(u) {\rm d} \xi_s^\alpha - \frac{1}{2} \int_0^T \phi^2_s(u) {\rm d} s } 
\rd u}{\int_t^\infty \rho_0(u)\, \re^{ \int_0^t \phi_s(u) {\rm d} \xi_s^\alpha - 
\frac{1}{2} \int_0^t \phi^2_s(u) {\rm d} s} {\rm d} u} \, H_T \right], 
\label{eq:36}
\end{eqnarray}
which is identical to the pricing formula under the ${\mathbb P}$ measure 
had $\{\alpha_t\}$ been identically zero in 
the first place, on account of the following observation. The price of the underlying 
asset at time $T$ can be expressed in the form 
\begin{eqnarray}
S_T = S_0 \exp\left( \int_0^T \left(r_s+\lambda_s\sigma_s-\half\sigma_s^2\right) 
\rd s + \int_0^T \sigma_s \rd \xi_s \right), 
\label{eq:37}
\end{eqnarray}
where $\{\sigma_t\}$ is the volatility of $\{S_t\}$. Now if the volatility of the 
martingale family $\{M_t(u)\}$ takes the form (\ref{eq:31}), then the risk premium 
can be expressed as 
\begin{eqnarray}
\lambda_t = \lambda_t^\alpha + \alpha_t, 
\label{eq:38}
\end{eqnarray}
where $\{\lambda_t^\alpha\}$ is the risk premium in the $\mathbb{P}^\alpha$ 
measure:
\begin{eqnarray}
\lambda_t^\alpha = -\frac{\int_t^\infty \rho_0(u) \phi_t(u) \re^{
\int_0^t \phi_s(u) {\rm d} \xi_s^\alpha - \frac{1}{2} \int_0^t \phi^2_s(u)
{\rm d}s} \rd u}{\int_t^{\infty} \rho_0(u) \re^{ \int_0^t \phi_s(u)
{\rm d}\xi_s^\alpha - \frac{1}{2} \int_0^t \phi^2_s(u) {\rm d} s} \rd u} . 
\label{eq:39}
\end{eqnarray}
Substituting (\ref{eq:35}) and (\ref{eq:38}) in (\ref{eq:37}), we obtain 
\begin{eqnarray}
S_T = S_0 \exp\left( \int_0^T \left(r_s+\lambda_s^\alpha\sigma_s
-\half\sigma_s^2\right) \rd s + \int_0^T \sigma_s \rd \xi_s^\alpha \right). 
\label{eq:40}
\end{eqnarray}
We thus find that the probability law of the random variable $S_T$, and hence 
$H_T$, under the ${\mathbb P}$-measure with $\alpha_t=0$, is the same as that 
under the ${\mathbb P}^{\alpha}$-measure with $\alpha_t\neq0$. It follows that 
any \textit{addition} of terms in the martingale volatility $\{v_t(u)\}$ that is 
independent of the parameter $u$ does not affect current price levels. 

The above result shows that the risk premium vector $\{\lambda_t\}$ can be 
determined from market prices of derivatives only up to an additive vectorial 
term $\{\alpha_t\}$. This freedom, however, is not arbitrary;  it can only 
arise from a constant (i.e. independent of the parameter $u$) addition to the 
volatility of the martingale family in the form of (\ref{eq:31}).

\section{Information-based interpretation}

The ambiguity in the determination of the risk premium can be interpreted from the 
viewpoint of information-based asset pricing theory of Brody \textit{et al}. (2007). 
In the information-based pricing framework one models the market filtration 
directly in the form of an information process concerning market factors relevant 
to the cash flows of a given asset. Our objective here, which extends the previous 
work of Brody and Friedman (2009), is to analyse the model (\ref{eq:16}) for the 
information process that determines the pricing kernel. 

The interpretation of the information process (\ref{eq:16}) is as follows. Market 
participants are concerned with the realised value of the random variable $X$, 
which, in a certain sense can be interpreted as the timing of a serious liquidity 
crisis. In reality, market participants observe price processes, or equivalently 
the underlying Brownian motion family $\{\xi_t\}$. As indicated above, under the 
physical ${\mathbb P}$-measure the random variables $X$ and $\xi_t$ are 
independent. However, market participants `perceive' information with certain 
risk adjustments characterised by the density martingale $\{N_t\}$ of (\ref{eq:25}). 
In this risk-adjusted measure, the path $\{\xi_t\}$ represents the aggregate of 
noisy information for the value of $X$ in the form of (\ref{eq:16}). The 
`signal' concerning the value of $X$, in particular, is revealed to the market 
through the structure function $\{v_t(u)\}$, which in turn determines the 
volatility structure of the pricing kernel, and hence the risk premium. 

Suppose that the structure function $\{v_t(u)\}$ takes the form (\ref{eq:31}), 
where $\{\alpha_t\}$ is independent of $X$. Then because (\ref{eq:16}) 
represents the information process for the random variable $X$, the constant 
$\{\alpha_t\}$ combines with the `noise' term $\{\beta_t\}$. In other words, 
the choice of $\{\alpha_t\}$ is entirely equivalent to the choice of noise; the 
Brownian noise is replaced by a drifted Brownian noise. This change of noise 
composition does not affect current asset prices, and therefore is not 
directly detectable from market data, even though asset-price drifts are 
modified, in general in an unidentifiable manner. Note that the point of view 
that the indeterminacy of the asset price drifts is caused by noise has been 
put forward heuristically by Black (1986); our observation thus formalises this 
argument more precisely. 

It is worth remarking briefly the observation made in Brody and Friedman (2009) 
concerning the form of the structure function $\{v_t(u)\}$ in the absence of the 
noise drift $\{\alpha_t\}$. Since small values of 
$X$ imply imminent liquidity crisis, in an ideal market the signal-to-noise ratio 
of the information process (\ref{eq:16}) should be large for small values of $X$, 
as compared to large values of $X$. In other words, under normal market 
conditions we expect the signal magnitude $|v_t(u)|$ be decreasing in $u$ for 
every $t$. Conversely, if $|v_t(u)|$ is increasing in $u$, then the excess rate of 
return above the short rate for discount bonds, i.e. the inner product of the risk 
premium and the discount bond volatility, is negative, yielding negative excess 
rate of return due to the inverted form of the structure function $\{v_t(u)\}$.

\section{Anomalous price behaviour}

The fact that current asset prices are unaffected by changes in the structure of 
the noise term does not imply that $\{\alpha_t\}$ can be ignored altogether. Indeed, 
(\ref{eq:38}) shows that the existence of such a component does shift the 
risk premium. Since the drift of an asset with volatility $\{\sigma_t\}$ is given in 
the ${\mathbb P}$-measure by $r_t+\lambda_t\sigma_t$, the noise-induced 
drift $\alpha_t\sigma_t$ can generate various anomalous price dynamics under 
the physical ${\mathbb P}$-measure. 

As an example, let us consider the case of an anomalous price growth, or a 
bubble. In the large vector space of asset volatilities, it is inevitable that 
volatility vectors form clusters consisting of different sectors or industries. This 
is because, by definition, a given sector of companies share analogous 
risk exposures. Now if an anomalous noise component $\{\alpha_t\}$ at some 
point in time emerges to point in the direction of one of these volatility clusters, 
then this can cause a sharp rise in the share prices of that sector. Since 
the noise vector $\{\alpha_t\}$ carries no real economic information, this can 
be identified as a bubble, where prices of a set of assets grow sharply, and 
independently of the `true' state of affairs, without seriously affecting price 
processes of other assets. Similarly, at a later time, the magnitude of 
$\{\alpha_t\}$ can diminish. In particular, more reliable information concerning 
the true state of affairs may be revealed, which in turn leads to an increase 
in the magnitudes of volatilities on the one hand, while on the other hand the 
risk premium vector can point in a direction such that the inner product 
$\lambda_t\sigma_t$ takes a large negative value; thus leading to a bubble 
`burst'. 

In the finance and economics literature, there exists a substantial work on the 
study of various aspects of financial bubbles (see, e.g., Camerer 1989 for an 
early review). It is important to note that our characterisation of a bubble is 
motivated by an information-based perspective. One commonly used definition 
of a bubble, on the other hand, is given by the difference between the current 
price and the expected discounted future cash flows in the risk-neutral 
measure (cf. Tirole 1985, Heston \textit{et al}. 2007). Under this definition, 
discounted asset prices in the risk-neutral measure can be modelled by use 
of strict local martingales (Cox and Hobson 2005, Jarrow \textit{et al}. 2007, 
2010), within the arbitrage-free pricing framework. 

While this formulation of a bubble leads to the unravelling of many interesting 
mathematical subtleties underlying fundamental theorems of asset pricing, from 
an information-theoretic viewpoint the plausibility of such a definition for a bubble 
seems questionable. In particular, a mathematical definition of a financial bubble 
that involves no reference to the ${\mathbb P}$ measure seems restrictive; a 
bubble, after all, is a phenomenon seen under the ${\mathbb P}$ measure.
The pricing kernel approach, on the other hand, is based on a stronger assumption 
that if $\{S_t\}$ represents the price process of a liquidly traded asset, then 
$\{\pi_tS_t\}$ must be a true ${\mathbb P}$-martingale. As such, the discounted 
${\mathbb Q}$-expectation of future asset price necessarily agrees with the 
current value, or else there are arbitrage opportunities. 

The conventional definition of a financial bubble in terms of the inequality 
$S_t>\pi_t^{-1}{\mathbb E}_t[\pi_T S_T]$ is sometimes justified heuristically by 
the fact that some traders, when they are under the impression that there is a 
bubble and thus traded prices are above the `fundamental' values, will 
nevertheless participate in the apparent bubble with the view that they can 
withdraw from their positions before the crunch (see, e.g., Camerer 1989 and 
references cited therein). This example and other similar ones are often used in 
support of the argument that some traders are willing to purchase stocks even 
when they 
know that the price level is above its fundamental value. The shortcomings in 
such an argument are that (a) the role of market filtration is not adequately taken 
into account; and that (b) the fact that such a stock purchase is equivalent to the 
purchase of an American option is overlooked. A more plausible characterisation 
of a bubble participation seems to be as follows. Given the information 
$\{{\mathcal F}_t\}$, a trader estimates that there is a bubble that will continue 
to grow for a while. Hence, subject to the filtration, the best estimate of the 
future cash flow for this trader, with a suitable risk-adjustment, is given by 
$\sup_{\tau}\pi_t^{-1}{\mathbb E}_t[\pi_\tau S_\tau]$, where $\tau$ is a stopping time 
when the stock is sold. If this expectation agrees to the current price level, then 
a transaction occurs. Conversely, it seems implausible that a transaction takes 
place if the best estimate by a rational trader of a discounted cash flow is lower 
than the current price level. 

The view we put forward here is that a bubble in an asset ought to be identified 
with an anomaly in the rate of return of that asset, and not with an anomaly in the 
price level itself. 
Here, a precise definition of an `anomaly' in the drift is essentially what we have 
described above, namely, the existence of an additive term in the volatility of 
the martingale family $\{M_t(u)\}$ that is constant in the parameter $u$. Based 
on this definition, it is admissible that price processes behave in a manner that 
does not always reflect what one might perceive as the true state of affairs, had 
one possessed better information concerning the true worth of the assets. Put 
the matter differently, decisions concerning transactions that ultimately lead to 
price dynamics are made in accordance with the unfolding of information. Since 
this information is necessarily noisy, the best filters chosen by market participants 
will inevitably deviate from true values of assets being priced. If the noise 
structure changes, then it is only reasonable that the dynamical aspects of these 
deviations will likewise change. In particular, the increment of the innovations 
representation---that characterises the arrival of `real' information over the 
interval $[t,t+\rd t]$---is 
given by 
\begin{eqnarray}
\rd W_t = \rd \xi_t - {\hat \phi}_t \rd t + \alpha_t \rd t,  
\label{eq:41}
\end{eqnarray}
where ${\hat\phi}_t={\mathbb E}_t^{\mathbb R}[\phi_t(X)]$, 
and this illustrates in which way the existence of a nonzero noise drift 
$\{\alpha_t\}$ affects the dynamics. 

Our characterisation of anomalous price dynamics is not confined to the 
consideration of financial bubbles. Again, in the large vector space of asset 
volatilities, it seems plausible that equity market volatilities and fixed-income 
volatilities generally lie on distinct subspaces. If the noise vector $\{\alpha_t\}$ 
has a tendency to lie in the direction of equity-volatility subspace, then this 
naturally leads to an excess growth in the equity market, explaining the 
phenomena of the so-called equity premium puzzle, where over time the 
rate of return associated with the equity market considerably exceeds that 
of the bond market (see, e.g., Kocherlakota 1996 for a review).

\section{Relation to the risk-neutral measure}

We have established the relation between the auxiliary probability measure 
${\mathbb R}$ and the physical measure ${\mathbb P}$. The relation between 
the latter and the risk-neutral measure ${\mathbb Q}$, on the other, involves the 
risk premium process $\{\lambda_t\}$. To recapitulate these two relations, we 
have 
\begin{eqnarray}
\rd W_t = \rd \xi_t - {\hat v}_t \rd t \quad {\rm and} \quad \rd W_t^* = \rd \xi_t 
+ \lambda_t \rd t,
\label{eq:42}
\end{eqnarray}
where 
\begin{eqnarray}
{\hat v}_t = \frac{\int_0^\infty \rho_0(u) v_t(u) M_t(u)\rd u}
{\int_0^{\infty} \rho_0(u) M_t(u) \rd u} \quad {\rm and} \quad 
\lambda_t = -\frac{\int_t^\infty \rho_0(u) v_t(u) M_t(u)\rd u}
{\int_t^{\infty} \rho_0(u) M_t(u) \rd u} ,
\label{eq:43}
\end{eqnarray}
and where we let $\{W_t^*\}$ denote the ${\mathbb Q}$-Brownian motion. By 
combining the two relations in (\ref{eq:42}) we deduce at once that the 
measure-change density martingale is given by 
\begin{eqnarray}
\left.\frac{\rd{\mathbb Q}}{\rd \mathbb{R}}\right|_{{\mathcal F}_t} = \exp\left(-
\int_0^t ({\hat v}_s+\lambda_s)\rd W_s -\half \int_0^t ({\hat v}_s+\lambda_s)^2 
\rd s \right),
\end{eqnarray}
which determines the general relation between ${\mathbb Q}$ and ${\mathbb R}$. 

As indicated above, a closer inspection on (\ref{eq:43}), however, shows that 
\begin{eqnarray}
{\hat v}_t = {\mathbb E}_t^{\mathbb R}[v_t(X)] \quad {\rm and} \quad 
\lambda_t = -{\mathbb E}_t^{\mathbb R}[v_t(X)|X>t] .
\label{eq:45}
\end{eqnarray}
In other words, under the restriction $X>t$, we have, conditionally, ${\hat v}_t+
\lambda_t=0$. Therefore, the auxiliary measure ${\mathbb R}$, whose existence 
is ensured by the lack of arbitrage and the existence of pricing kernel, can be 
interpreted as an extension of the risk-neutral measure. Conversely, by 
restricting to the event $X>t$, we can think of ${\mathbb R}$ indeed as the 
risk-adjusted measure.

\section{Stochastic volatility} 

So far we have analysed the case for which $\{v_t(x)\}$ is a deterministic function 
of time. The fact that the volatility structure of the martingale family $\{M_t(x)\}$ is 
deterministic, however, does not imply deterministic volatilities for asset prices. 
On the contrary, even for an elementary discount bond, the associated volatility 
process is highly stochastic. Hence when we speak about a `stochastic volatility' 
we have in mind the volatility for the martingale family $\{M_t(x)\}$, whereas the 
stochasticity for asset prices is presumed. 

From the viewpoint of practical implementation, it probably suffices to restrict 
attention to deterministic volatility structures, since deterministic volatilities for 
$\{M_t(x)\}$ give rise to a range of sophisticated stochastic volatility models for 
asset prices. Indeed, it is shown in Brody \textit{et al}. (2011) that even in the 
very restricted case of a single factor model with the time-independent volatility 
$v_t(x)=\re^{-\sigma x}$ that depends only on one model parameter $\sigma$, it is 
possible to calibrate caplet prices across different maturities reasonably accurately. 

It is nevertheless of interest to enquire whether the auxiliary information process 
exists in the more general context of stochastic volatilities. For this purpose, let 
us begin by considering the case where $\{v_t(x)\}$ admits the decomposition 
(\ref{eq:31}) and where $\{\phi_t(x)\}$ is deterministic and $\{\alpha_t\}$ is chosen 
such that the noise term $n_t\equiv\int_0^t\alpha_s\rd s+\beta_t$ is an 
$\{{\mathcal F}_t^\beta\}$-measurable Gaussian process. Then an application of 
the martingale representation theorem shows that $\{n_t\}$ admits a 
decomposition of the form 
\begin{eqnarray}
n_t = \int_0^t b_s \rd s + \int_0^t \gamma_s \rd \beta_s,
\end{eqnarray}
where $\{b_s\}$ and $\{\gamma_s\}$ are deterministic. A short calculation then 
shows that an auxiliary information process 
\begin{eqnarray}
\xi_t = \int_0^t \phi_s(X) \rd s + n_t
\end{eqnarray}
in the ${\mathbb R}$ measure indeed exists, with the property that the scaled 
information process $\int_0^t\gamma_s^{-1}\rd\xi_s$ determines the market 
Brownian motion and that $\{b_t\}$ plays the role similar to that of a 
deterministic $\{\alpha_t\}$ in the previous analysis, and hence is not determinable 
form current market prices. 

The foregoing example shows how one can model the random rise and fall of 
anomalous price dynamics. More generally, the structure function $\{v_t(x)\}$ can 
depend in a general way on the history of the information process up to time $t$. 
In this case, we obtain a generic stochastic volatility model for the martingale 
family. Provided that the structure function is sufficiently well behaved so that 
relevant stochastic integrals exist, the auxiliary information process can be seen 
to exist in the ${\mathbb R}$ measure. To illustrate this, consider an elementary 
`toy model' for which information process takes the form of an Ornstein-Uhlenbeck 
process: 
\begin{eqnarray}
\xi_t = \re^{\sigma X t}\int_0^t \re^{-\sigma Xs} \rd \beta_s,
\end{eqnarray}
where $\sigma$ is a parameter, and $X$ and $\{\beta_t\}$ are independent. Such 
an information process corresponds to a stochastic volatility model for which the 
volatility process is given by a linear function of the ${\mathbb P}$-Brownian 
motion: $v_t(x)=\sigma x \xi_t$.

\section{Discussion} 

The main results of the paper are as follows: We have derived the existence of an 
auxiliary filtering problem underlying arbitrage-free modelling of the pricing kernel; 
the solution of which determines the volatility structure of the 
positive martingale family $\{M_t(u)\}$ appearing in the Flesaker-Hughston 
representation for the pricing kernel. We have demonstrated that the structure of 
the ambient information process fully characterises the risk premium process 
$\{\lambda_t\}$. We have shown, under the Brownian-filtration setup, that 
$\{\lambda_t\}$ admits a canonical decomposition into two terms in an additive 
manner; the systematic term that can be calibrated from current market data for 
derivative prices, and the idiosyncratic term that cannot be estimated (unless, of 
course, one can estimate drift processes of risky assets), and thus can be 
identified as noise.

It is worth emphasising that these results hold irrespective of our choice of 
interpretation. Nevertheless, our characterisation of anomalous price dynamics 
seems sufficiently compelling, for, such phenomena are ultimately observed 
under the physical measure ${\mathbb P}$. One might ask what causes the 
evolution of the noise drift $\{\alpha_t\}$. This is an interesting 
econometric question that, however, goes beyond the scope of the present 
investigation. It 
suffices to remark that the random variable $X$ that constitutes the signal 
component of the ambient information process has units of time, and thus is 
ultimately linked to the term structure of financial markets. One possible 
explanation of the excess equity premium therefore is that fixed-income market 
intrinsically embodies more information concerning the term structure as 
compared to the equity market, and this imbalance is manifested in the form of 
an additional drift in the noise component pointing generally towards the 
direction of equity volatility vectors.

\vskip 10pt 

\noindent The authors thank Mark Davis, Robyn Friedman, Matheus Grasselli, 
Lane Hughston, Andrea Macrina, and Bernhard Meister for comments and 
stimulating discussions.

\end{document}